# AUTOMATIC PREDICTION OF SMALL GROUP PERFORMANCE IN INFORMATION SHARING TASKS

Wen Dong, Bruno Lepri, Alex (Sandy) Pentland

MIT Media Lab
Ames Street
Cambridge, MA, 02124, USA
e-mail: {wdong,pentland,brulepri}@media.mit.edu

## ABSTRACT

In this paper, we describe a novel approach, based on Markov jump processes, to model small group conversational dynamics and to predict small group performance. More precisely, we estimate conversational events such as turn taking, backchannels, turn-transitions at the micro-level (1 minute windows) and then we bridge the micro-level behavior and the macro-level performance. We tested our approach with a cooperative task, the Information Sharing task, and we verified the relevance of micro-level interaction dynamics in determining a good group performance (e.g. higher speaking turns rate and more balanced participation among group members).

## INTRODUCTION

Nowadays, management, scientific research, politics and a lot of other activities are accomplished by groups. For this reason, it is becoming even more important to understand the determinants of group performance. The research area of organizational behavior has proposed and tested methods to improve the effectiveness of group collaboration and to deal with the problem of group sub-optimality, groups tend to perform better than individuals but not as well as they could [Kerr and Tindale, 2004]. In particular, group dynamics have been one of the focuses as it is a key factor affecting the performance and the satisfaction of the group [Shaw, 1976].

For instance, Hall and Watson [Hall and Watson, 1970] demonstrated that the performance of a group is noticeably affected by the understandings from its members on what is a productive group process, and that the group performance could be improved by just instructing the group members to be more participative and engaged in the conversation. According to them [Hall and Watson, 1970], a more productive group is more likely to generate group answers that are better than the individual answers by reconciling the differences among its members with win-win strategies and through 'aha' experiences.

Wilson et al. [Wilson et al., 2004] observed several tens of group processes in solving two versions of the 20-questions game. They noted that (1) groups solve significantly larger proportions of the games than individuals, (2) the questions asked by groups work increasingly better than those asked by individuals as a game proceeds and becomes harder, (3) a pair of strangers generate more (unique) ideas — that are compatible with a given list of yes/no questions and their answers — than a pair of friends, and a pair of friends generate more ideas than two individuals working alone. Many issues related to the lacking of participation, such as social-loafing and production-blocking, have been discussed by various researchers [Karau and Williams, 1993].

A very recent and interesting study has shown that groups perform better on tasks if the members have strong social skills and if the conversation reflects more group members' ideas [Woolley et al., 2010]. The tasks could range widely from brainstorming to quantitative analysis to negotiation and are drawn from all the quadrants of the McGrath Task Circumplex [McGrath, 1984], a well-established taxonomy of group tasks based on the coordination processes they require. The major findings were that group performance is not related to the average or maximum of the members' performances but it is correlated with average social sensitivity of group and with the equality in distribution of turn taking. This and many previous findings support the speculation that certain aspects of the interactions among the members are important to group performance and are independent of specific tasks.

In this paper, we propose and discuss our approach to relate the microcosmic interaction patterns among the group members to the group performance. We propose to use the Markov jump process model, an extension of Markov chains when time is considered to be continuous instead of discrete, in order to capture how the microcosmic interaction patterns will generate the macrocosmic interaction statistics such as the equal participation among the group members and the engagement, which in turn will have consequences on group performances. In particular, we focus on modeling the form of the interaction and conversation in small group meetings. Our proposed Markov jump process models the turn taking in small group conversations following the turn taking *systematics* proposed by Sacks et al. [Sacks et al., 1974]. Roughly speaking, the turn taking *systematics* consists of turn-constructional features for determining where transitions will be relevant, two types of turn-allocational techniques (current speaker selects the next one and self-selection) for determining how a next turn will be allocated, and a set of practices for employing the turn-allocational techniques by reference to transition-relevance places. In sum, in this model the current speaker selects the next speaker, the next speaker self-selects himself or the current speaker continues at transition places. Sacks et al. used this simple systematic to explain how the conversations are locally managed, party administrated, interactively controlled and sensitive to recipient design.

We believe that the automatic prediction of small group performance through tracking the turn taking behavior from signals such as audio variance, motion, and who-faces-whom has the following advantages: (i) it is computationally cheap and power efficient; (ii) it combines multiple types of sensor data in a unified framework and achieves better performance by related different types of signals; (iii) it tells us how microcosmic interaction can have macrocosmic consequence on performance.

The rest of the paper is organized in the following way. In the next section, we describe previous work on automatic recognition of small group performance and outcomes. Then, we describe our Markov jump process framework to capture the structure-performance and the microcosmic-macrocosmic relationships about small group dynamics. In the fourth section, we describe a group cooperative task, the Information Sharing task, tracked using socio-metric badges. After that, we show how our approach, the Markov jump process framework, is able to track and to predict the small group performance. Finally, we draw some conclusions about this study.

## PREVIOUS AND RELATED WORKS

Recent works in the automatic behavior analysis started to model some relevant dynamics for the small group performance and to predict the performance obtained in different tasks using simple acoustic and visual non verbal features.

Lepri et al. [Lepri et al., 2009] addressed the possibility of predicting the individual performances of subjects involved in a small group problem solving task (Mission Survival) by means of short sequences of non verbal behavior, so called "thin slices". Dong et al. [Dong et al., 2009] analyzed the relationship among the brainstorming performance (e.g. the number of ideas generated) and the decision making performance and then they analyzed the relationship among the behavioral interactions between the meeting participants and their individual performances.

Kim et al [Kim et al. 2008] developed a real-time portable system, Meeting Mediator, able to detect social interactions and to provide a persuasive feedback to enhance the group collaboration and cooperation. In this system, the social interactions are detected using Sociometric badges [Olguín Olguín et al., 2009] and are visualized on mobile phones to promote behavioral change. Particularly in distributed collaborations, MM attempts to bridge the gap among the distributed groups by detecting and communicating social signals. In a study on brainstorming and problem solving meetings, MM had a significant effect on overlapping speaking time and interactivity level without distracting the subjects. The Sociometric badges were also able to detect dominant players in the group and measure their influence on other participants. Most interestingly, in groups with one or more dominant people, MM effectively reduced the dynamical difference between co-located and distributed collaboration as well as the behavioral difference between dominant and non-dominant people.

In [Dong and Pentland, 2010], the authors discuss how group performance is related to several heuristics about small group dynamics enacted in performing several typical tasks (e.g., brainstorming, shopping tasks, problem solving, judgment task). The authors also proposed a new stochastic model to learn the dynamics of small group interactions and showed how is possible predicting significantly (R2 value up to 40%) the group performance using non-linguistic vocal statistics such as the number of clauses, the speaking speed, the number of vowels, the speaking turn length, the overlapping speaking, and so on.

Finally, Hung and Gatica Perez [Hung and Gatica Perez, 2010] investigated systematically automatically extracted acoustic and visual features that can be used to measure cohesion levels in small group meetings. In this study, the more predictive feature is an acoustic cue, which accumulated the total pause time between each individual's turns during a meeting segment.

## MODELING CONVERSATIONAL DYNAMICS

We use Markov jump process, an extension of Markov chains when time is considered to be continuous instead of discrete, to estimate conversational turns by using the following multimodal cues: (i) speech variance, (ii) body movement variance collected using a 3 axes accelerometer, (iii) who faces whom by means of infrared scanning The rationale of using not only speech variance but also body movement variance and information about face-to-face interactions is based on some background literature. In Kendon [Kendon, 1967] was showed that the addresser-addressee pair can be easily determined by who faces whom. Then, Harrigan [Harrigan, 1985] found that the amount of listeners' bodily activation is correlated with the speaking activity of the speaker and have a relevant impact on the conversational dynamics (e.g. turn-taking).

Markov jump process is likely to output that a person is speaking if his recorded audio intensity is greater than an estimated threshold, and we carefully adjust the thresholds of the persons in a group with an optimization algorithm so that the turn-taking structure is maximally satisfied. The audio intensity for an individual in a group discussion is assumed to be a linear combination of the audio intensities of all individuals in the discussion, and the intended individual has more contribution to the intensity. In our framework, we define a *speaking turn* as one continuous segment, not less than 1.5 sec., where a participant starts and ends her/his speech. Then, we modeled the following aspects of the turn-taking structure: (i) *taking the turn*: if nobody is taking the turn, then somebody should take the turn; (ii) *backchannel* [Yngve, 1970]: we define backchannel as the situation where a subject Y speaks after a subject X for less than 1 sec. (e.g. "yes" or "uh-huh"); (iii) *speaker transitions*: if somebody is ending the turn, then she/he will transfer to another person. Roughly speaking, we have a *speaker transition* instead of a *taking the turn* when a speaking turn of a subject Y follows in systematic way the speaking turn of a subject X ; (iv) *turn competition*: if two persons competing for turn, then one person will win. We define a *turn competition* as a situation in which 2 subjects are speaking at the same time and one ends before the other.

Specifically, the conversational state consists of whether speakers have turns. The conversational state at time $t$ is expressed as a **state vector** $x(t) = \left( x^{(1)}(t)? = 0, x^{(1)}(t)? = 1, \vdots, \cdots, \vdots, x^{(C)}(t)? = 0, x^{(C)}(t)? = 1 \right)'$ where there are $C$ speakers and $x^{(c)}$ is either 0 or 1 representing speaker c is not speaking or speaking. In general, elements of $x(t)$ can contain any value besides Boolean values, such as the number of chemicals in chemistry, the number of species in ecology and the price of an asset in economy.

Conversational state $x(t)$ is changed by different events $(1, \cdots, v)$, and it also determines the rates $\hbar_v(x(t))$ at which different events will happen. We use **event vector** to describe the number of different events happening in a time window: $r = (r_1, \cdots r_v)'$ where $r_i$ is the number of events of type $i$. We denote an event by a "reaction" $\sum_i \alpha_i x_i \rightarrow \sum_j \beta_j y_j$ where $\alpha_i$ number of reactant $x_i$ has been consumed and $\beta_j$ number of product $x_j$ has been generated. In our model of conversational dynamics, an event moves turn-taking status, and $\alpha_i$, $\beta_j$ are all one. We care about 4 types of events in our modeling: taking a turn, yielding a turn, transferring a turn and speaking in a back channel. We used Bayesian priors to bias the event rates towards reasonable values and tune the hyper-parameters of the priors manually. We considered 36 events in our analysis of four-person conversations: 4 different rates for the four persons to take a turn when nobody is currently taking the turn, 4 rates to yield a turn, $4 \times 4$ rates to transfer turn, 4 rates to speak in a back channel, 4 rates to seize a turn when another one is having the turn, and 4 rates to yield a turn when two or more persons try to take their turns simultaneously.

We use matrix algebra to express how events change conversational state. To this end we define the **reaction matrix** $A$ as a $C \times v$ matrix where $C$ is the length of the state vector $x(t)$ and $v$ is the number of reactions. An element at column $k$ and row $j$ represents the amount added to state $x_j(t)$ if reaction $k$ happens. In our modeling of conversational dynamics, entries of $A$ are either $+1$ or $-1$ representing moving into a state or moving out of a state. For example, in the following equation, the first three columns of $A$ represent speaker 1 starts to speak, speaker 2 transfers turn to speaker 3, and speaker 4 stops speaking. The column vector $r$ means a speaker-transition event has happened. If we multiply A by r, we get an update the state matrix.

$$A \cdot r = \begin{pmatrix} -1 & & & \\ 1 & & & \\ & 1 & & \\ & -1 & & \cdots \\ & -1 & & \cdots \\ & 1 & & \\ & & 1 & \\ & & -1 & \end{pmatrix} \cdot \begin{pmatrix} 0 \\ 1 \\ \vdots \\ 0 \end{pmatrix} = \begin{pmatrix} 1 \\ -1 \\ -1 \\ 1 \end{pmatrix}$$

$$= \Delta x$$

Let $r(t_i)$ be the event vector representing the numbers of different events happening between $t_i$ and $t_{i+1}$. In the ideal situation $r_{v_i}(t_i) = 1$ and the other elements of $r(t_i)$ are 0 because only event $v_i$ happened during the period. The system states starting from $x(t_0)$ and corresponding to the sequence of events are updated according to $x(t_{i+1}) = x(t_i) + A \cdot r(t_i)$.

In order to derive the inference algorithm for estimating turn-taking dynamics from noisy sensor data, we begin with the ideal situation that we know all events $(t_i, v_i)$, where $i = 1, \cdots, n$, $0 = t_0 < t_1 < \cdots < t_n = T$ and $v_i \in \{1, \cdots, v\}$. The probability for this sequence of events to happen is

$$\mathrm{P}(v, x) = \prod_i h_{v_i}(x_t) \cdot \exp\left(-\sum_i h_{v_i}(x_t) \cdot t_{i+1} - t_i\right)$$

In reality we only have discrete time observations $y(n \cdot \Delta t)$ such as audio variance, body movement variance and detection of face-to-face configuration, and we want to infer from these discrete time observations how many, when and what events happened between these observations. The inference algorithm becomes non-trivial when the time interval between two consecutive observations becomes large, when we have missing data, and when we have data that are incompatible with the model. However it is possible to construct exact MCMC algorithms for inference based on discrete time observations 0, and it is possible to make inference with mean field approximation and variational method 0.

We introduced the following approximations to make the inference of turn-taking dynamics conceptually much simpler. Our first approximation is that turn-taking events only happen at the times of observation, and this approximation introduces 0.05 second error in the event times. Our second approximation is that at most one event can happen between two consecutive observations or 0.1 second. Our third approximation is that the observations for inferring turn-taking state have joint Gaussian distributions conditioned on turn-taking state.

Thus the probability of a sequence of latent events $v(t)$, together with the corresponding latent states $x(t)$ and observations $y(t)$ is

$$P(v, x, y) = \prod_{t,c} P\left(y^{(c)}(t) \mid x^{(c)}(t)\right) \cdot P(x(0)) \prod_t P\left(v(x(t))\right),$$

$$P(v(x(t)) = 0) = \exp\left(-\sum_j h_j(x_t)\right),$$

$$P(v(x(t)) = i) = \frac{h_i(x(t))}{\sum_j h_j(x(t))}\left(1 - \exp\left(-\sum_j h_j(x(t))\right)\right),$$

$$P\left(y^{(c)}(t) \mid x^{(c)}(t)\right) = N\left(\mu_{x^{(c)}(t)}, \Sigma_{x^{(c)}(t)}\right).$$

We use Gibbs sampling to infer latent states and parameters:

$$v(x(t)) \mid \mu, \Sigma \sim Categorical\left(\left\{P(v(x(t)) = i P_y t x t P_v v x t + 1 : i = 0, \cdots, v\right\}\right)$$

$$h_i \sim Dirichlet\left(h + \sum_t \delta_{v(x(t)),i}\right).$$

$$\Sigma \mid \kappa_0, \mu_0, \nu_0, \Psi_0 \sim W^{-1}\left(T + \kappa_0, \Psi_0 + \sum_t (y_t - yyt - yT + nv0n + v0y - \mu0y - \mu0T,\right.$$

$$\mu \mid \Sigma, \nu_0, \mu_0 \sim N\left(\frac{n\bar{x} + \nu_0\mu_0}{T + \nu_0}, \frac{\Sigma}{n + \nu_0}\right).$$

## INFORMATION SHARING TASK

We tested our Markov jump process framework on a dataset collected to examine the relationship between the communication patterns and the group performance during a cooperative task. This dataset was collected from 50 groups of four members each, for a total of 200 participants. Each participant wore a Sociometric badge [Olguín Olguín, 2009], a wearable electronic device with multiple sensors (e.g. microphone, infrared, accelerometer) able to detect face-to-face interactions, physical proximity, body movement data and speech features. Regarding the speech, due to privacy concerns the badges do not collect content of the speech or any other feature that may identify the speaker. The Figure 2 shows an example of participants wearing sociometric badges.

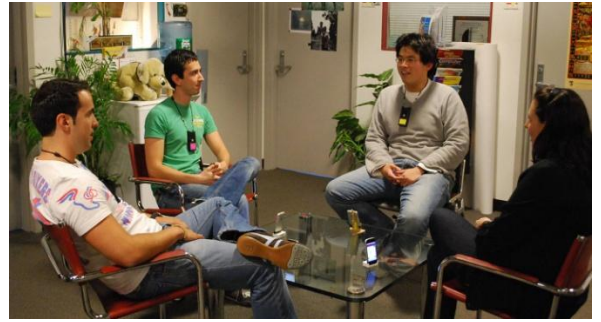

*Figure 1: Example of an interacting group wearing sociometric badges around the neck.*

We used a variation of an hidden-profile task [Mesmer-Magnus and Church, 2009], which measures how well the group members shares the information. The group cannot successfully perform the task unless the participants pool all the information that the members individually hold. It means that people must reason about what fact from their private information will lead to the greatest accuracy gain in the public information base.

We apply the format of a 20-questions game to the hidden-profile task to have a quantitative measure on how well the information is shared among members. In this way, the possible answer space is strictly confined and equally divided among the members. More, information from all members is required for the group to generate more efficient questions.

Each member was given a sheet of paper with a list of 10 people (possible answers) along with three attributes of their personal information, which were height, weight, and a test score. Each member's sheet had a non-overlapping set of possible answers; hence there were 40 possible answers in total among the four members. The goal of the game was to correctly guess the one person that the experimenter is thinking out of the 40 possible answers. Groups discussed to generate a yes-or-no question, which narrows down their answer space. When a question is generated, the experimenter answers the question by either a yes or a no, after which the groups continue their discussion to generate the next question based on the answer that they heard. This process was repeated until the group came to the correct answer. The number of yes-or-no questions that the group needed to come to the correct answer was the inverse-measure of the group performance. One of the participants was chosen as a task coordinator immediately before the task starts. The task coordinator's role was to be the channel of communication between the subjects and experimenter. For the first question, up to 4 minutes were provided for question generation, and 2 minutes were provided for the following questions.

The optimal strategy to quickly arrive to the answer was to ask yes-or-no questions that would narrow down the possible answer space in half. This strategy would guarantee that groups could come to the correct conclusion within log2 (40) = 5.32 number of questions. This strategy was informed to all participants before the task started. Hence, all groups aimed to generate a question that divides the answer space into half. In order to generate this optimal question, they needed to communicate verbally to correctly understand the distribution of the three attributes of the possible answers. If one or more members withheld the information that they had, the group would have a biased understanding of the answer space, which resulted in asking questions that did not halve the possible answer space. Examples were given to make sure each participant had full understanding the task and the optimum strategy. Once all participants understood the study, the lists of possible answers were given to the participants. This list was different for each participant and they were non-overlapping. The participants were later allowed to talk about their list, but they were never allowed to show this list to other members. All members were given a short time (1-2 minutes) to look over the list individually before starting the conversation with other members.

Groups were given up to 4 minutes to generate their starting question, and additional 2 minutes per following questions. After the group came up with a question, the task-coordinator of the group would raise his/her hand to notify their question to the experimenter. The experimenter recorded the start and end times of each question generating phase, as well as the question that the group generated.

## AUTOMATIC PREDICTION OF GROUP PERFORMANCE

We extracted group dynamics from body movement variation (through a 3-dimensional accelerometer), audio signal variation, and who faces whom (through directional infrared detection) from the Sociometric badges. We aligned the sensor data from different badges using the time-stamped Bluetooth messages going from badge to badge. We further aligned the sensor data by aligning the time points with greater than 90 percentile audio amplitudes on different badges. The resulting signals are aligned within 0.02-second error. We initialize our turn estimation algorithm by locating audio segments with greater than 90 percentile as potential pitched segments, apply mixture of Gaussian distributions model with two states on inter-"pitch" gaps to find potential turn breakings (as gaps approximately greater than .7 second), using body movement variation of the others and the previous infrared detection as hints.

The Information Sharing task has some specific group behavioral dynamics: (i) short turn lengths (1 second on average), (ii) fast speaker transitions (40 to 80 turns and back channel instances per minute) and a significant amount of parallel speaking to facilitate information gathering.

The following claims on performance-interaction relationships are supported by one-sided Wilcoxon signed ranked[1] tests [Wilcoxon, 1945] at a significance level 0.01. The group performance of the Information Sharing task is measured in terms of number of questions asked. Theoretically a group needs no more than 6 questions to solve the task if its members have complete information of one another, and in reality the groups asked from 5 to 8 questions to get the answers. During the task, the conversational events decrease the number of possible answers by different fractions and lead to the problem solution when only one answer left. The effect of a question is the combined effect of the conversational events between this question and the previous one. So, we used a duration model, the proportional hazard model [Breslow, 1975], to correlate the fraction of possible answers deleted after each question with the rates of the different conversational events (transferring a turn, taking a turn when it is available, speaking in back channel and interruption). Survival analysis is concerned with modeling life span $S(t) = \Pr(T > t)$ in different application areas, and two important functions in survival analysis are the hazard function $\lambda(t)$ and the cumulative hazard function $\Lambda(t)$: $\lambda(t)dt = \Pr(t \leq T \leq T + dt | t \leq T) = d \log S(t) =$

---

[1] The Wilcoxon signed-rank test is a non-parametric statistical hypotheses test used to compare two related samples or repeated measurements on a single sample in order to assess whether their population means differ

$\Lambda(t)$. Proportional hazard model assumes that the hazard function is proportional to the covariates $\lambda(t|X) = \lambda_0(t) + \sum \beta_p X_p$ where $X_p$ are covariates.

In the 20-questions task: each competition for turns and each turn transferring respectively increase hazard rate by 1e-3 and 1e-4 $(p < 0.001)$, both leading to more questions and worse performance; Each turn taking and each instance of speaking in back channel respectively increase hazard rate by 2.5e-4 $(p < 0.001)$ and 1e-5 $(p < 0.01)$. Interruption in this experiment normally happens at the beginning of a turn and indicates a failure in getting a turn. Turn transferring in this experiment reflects the fact that a subject tends to speak with another subject in the same place in collocated settings. Survival analysis explains 18% variance of 20-questions performance from conversational events.

We can also construct a look-up table to explain the performance (good or bad) from group discussion dynamics such as the averaged number of speaking turns, the averaged number of turn competitions, the averaged number of backchannels, and the averaged number of speaking turns played by different subjects in a group discussion of 1 minute. To clarify the difference among number of speaking turns and number of speaking turns played by different subject, let we make the following example: during a group discussion meeting of 4 members (X, Y, Z, W) in a 1 minute time window first X have a turn, then Y have a turn and finally X take again the turn. In this scenario, the number of speaking turns is equal to 3 and the number of speaking turns played by different participants is equal to 2.

| Performance percentile | Turn taking | Turn competitions | Back-channel | Turn taking by different members |
|---|---|---|---|---|
| 25% | 30 | 2 | 10 | 25 |
| 50% | 40 | 4 | 18 | 30 |
| 75% | **50** | 5 | 15 | **35** |

*Table 1: Relationship among performance and group discussion dynamics in 20-questions task.*

Table 1 is computed first by estimating the rates of conversational events at different performance percentiles, and then repeatedly simulating conversational event sequences from the estimated event rates using Markov jump process, and finally counting the statistics from the simulated sequences.

As shown in Table 1, our results suggest that a more active group discussion (higher averaged number of group speaking turns) and a more balanced or egalitarian discussion participation among the members (higher number of speaking turns played by different members) have a better performance as outcome.

We have also reconstructed the remaining items immediately before each of the 324 questions asked by the groups. We identify a situation with unbalanced information-sharing among the members focusing on 74 questions where the distribution of the remaining items has a log2-entropy below 1 indicating that the remaining items were mostly at the hands of two people out of four. Our goal was to verify if there is a relationship between unbalanced information among the group members and unbalanced participation in the group discussion. In order to do this, we verify the feasibility of predicting the fraction of speaking time of a given subject X from the fraction of the remaining items of the same subject and the opposite. We are able to predict it within 40% variance $(p < 0.01)$ using a linear regression model over all cases.

Then, we focus our analysis on the bad performance. We identified 17 "bad" questions that would lead to 30% more remaining items in the worse case. Our goal was to test if there is a relationship between the low conversational participation and the "bad" questions. So, we ran a linear regression where the independent variable is the fraction of subject X speaking time and the dependent variable is the difference between the number of eliminated items by X and the number of items that X would eliminate after a "good" question. We are able to predict the fraction of speaking time of a subject from the distribution of remaining items among the subjects between the asked question and the should-be question with a 45% variance.

Finally, using only the previous turn-taking behavior of a subject X we can explains 40% variance of his future turn-taking behavior (e.g. if a person speaks 3 turns per minute for the choosing first question it is likely that she/he will speak around 3 turns for choosing the next question). Instead, using also information about the discussion dynamics of the other members we can explain 47% variance. In order to compare the performance of these 2 model, the first predict the turn statistics of subject X looking only at the previous turn statistics of Subject X (subject-based model) and the second predicts the turn statistics of subject X looking also at the turn statistics of all the other subjects (group dynamics-based model), we ran an analysis of variance (ANOVA) test. The results of our test support the better predictive power of the group dynamics-based model at significance level $p < 0.01$.

## CONCLUSIONS

The aim of this paper was to investigate and to test a novel approach to reason about group performance through modeling and sensing small group interaction dynamics. We propose a Markov jump process framework to estimate conversational events such as turn taking, turn transitions, competition to take the turn and back channel looking at 1 minute windows, bridge micro-level non verbal behaviors and macro-level performance.

We test our model on a cooperative task dataset collected in a lab setting using wearable devices able to extract speech features (e.g. speech variance), to capture face-to-face interactions by means of infrared sensors, and subjects' body motions by means of 3-axes accelerometers.

One pillar of our work is using only simple non-verbal multimodal cues (speech variance, body movement variance, and information about who is face whom) not related to the semantic contents of the interactions.

The main findings of our paper are the following: (i) the groups with a good performance have a more active discussion (more speaking turns) and a more balanced participation among the members (more speaking turns are played by different subjects) and (ii) there is a relationship between unbalanced information sharing and unbalanced involvement in discussion dynamics.

On the practical side, our results are important steps towards automatic systems able to analyze, assist and modify small group dynamics in order to provide various kinds of support to dysfunctional teams, from facilitation to training sessions addressing both the individuals and the group as a whole. On a more theoretical side, our work emphasizes the relevance of micro-level interaction dynamics in determining if the group performance will be good or bad.

## AKNOWLEDGMENTS

Bruno Lepri's research is funded by PERSI project inside the Marie Curie COFUND-7th Framework

## REFERENCES

Boersma, P. and Weenink, D. (2011). *Praat: doing phonetics by computer* [Computer program]. Version 5.2.25, retrieved 11 May 2011 from http://www.praat.org/

Boys, R. J., Wilkinson, D. J. and Kirkwood, T. B. L. Bayesian inference for a discretely observed stochastic kinetic model. (2008). *Statistics and Computing*. 18(2) 125-135.

Breslow, N. E. (1975). Analysis of Survival Data under the Proportional Hazards Model. *International Statistical Review / Revue Internationale de Statistique* 43(1) 45–57.

Dong, W., Kim, T., and Pentland, A. (2009). A quantitative analysis of collective creativity in playing 20-questions games. *ACM Cognition and Creativity*.

Dong, W. and Pentland, A. (2010). Quantifying group problem solving using social signal analysis. *ACM ICMI 2010*.

Hall, J. W. and Watson, W. H. (1970). The Effects of a Normative Intervention on Group Decision-Making Performance. *Human Relations*. 23(4) 299-317.

Harrigan, J.A. Listeners' Body Movements and Speaking Turns. (1985). *Communication Research* 12(2) 233-250.

Hung, H and Gatica-Perez, D. Estimating Cohesion in Small Groups using Audio-Visual Nonverbal Behavior. (2010). *IEEE Transactions on Multimedia*. 12(6) 563-575.

Karau, S.J, and Williams, K.D. (1993). Social loafing: A meta-analytic review and theoretical integration. *Journal of Personality and Social Psychology*. 65 681–706.

Kendon, A. (1967). Some Functions of Gaze Direction in Social Interaction. *Acta Psychologica*. 26 22–63.

Kerr, N.L. and Tindale, R.S. (2004). Group performance and decision making. *Annual Review of Psychology*. 55 623-55.

Kim, T., Chang, A., Holland, L. and Pentland, A. (2008). Meeting mediator: enhancing group collaboration using sociometric feedback. *ACM CSCW*. 457-466.

Lepri, B., Mana, N., Cappelletti, A., Pianesi, F. (2009). Automatic Prediction of Individual Performance from Thin Slices of Behavior. *ACM Multimedia*.

McGrath, J. E. (1984). Groups: Interaction and Performance. Prentice-Hall, Englewood Cliffs, NJ.

Mesmer-Magnus, J. R. and Church, L. A. (2009). Information Sharing and Team Performance: A Meta-Analysis. *Journal of Applied Psychology*. 94(2) 535-546.

Olguín Olguín, D., Waber, B., Kim, T., Mohan, A., Ara, K. and Pentland, P. (2009). Sensible Organizations: Technology and Methodology for Automatically Measuring Organizational Behavior. *IEEE Transactions on Systems, Man, and Cybernetics-Part B: Cybernetics*. Vol. 39, No. 1, pp. 43-55.

Opper. M. and Sanguinetti. G. (2007). Variational inference for Markov jump processes. *NIPS 2007*.

Sacks, H., Schegloff, E. A. and Jefferson, G. (1974). A simplest systematics for the organization of turn-taking for conversation. *Language*. 50(4) 696-735.

Shaw, M. E. (1976). Group dynamics: the psychology of small group behavior. New York, McGraw-Hill.

Wilcoxon, F. (1945). Individual comparisons by ranking methods. *Biometrics Bulletin*. 1(6): 80–83.

Wilson, D. S., Timmel, J. J. and Miller, R. R. (2004). Cognitive cooperation: When the going gets tough, think as a group. *Human Nature*, 15(3):1–15.

Woolley, A. W., Chabris, C. F., Pentland, A., Hashmi, N. and Malone, T. W. (2010). Evidence for a collective intelligence factor in the performance of human groups. *Science*. 330(6004) 686-688.

Yngve, V. On getting a word in edgewise. (1970). Papers from the sixth regional meeting of the Chicago Linguistic Society, pages 567–77, Chicago: Chicago Linguistic Society.